\documentclass[%
 reprint,
 amsmath,amssymb,
 aps,
]{revtex4-1}

\usepackage{float}

\usepackage{xspace}
\usepackage{graphicx}% Include figure files
\usepackage{dcolumn}% Align table columns on decimal point
\usepackage{bm}
\usepackage[version=4]{mhchem}
\usepackage[utf8]{inputenc}
\usepackage[T1]{fontenc}
\usepackage{mathptmx}
\usepackage{siunitx}

\newcommand{\hfo}{HfO$_\text{2}$\xspace}
\newcommand{\zro}{ZrO$_\text{2}$\xspace}
\newcommand{\hzo}{Hf$_{1-x}$Zr$_x$O$_2$\xspace}

\DeclareSIUnit\fu{f.u.}
\begin{document}

%\preprint{AIP/123-QED}

%\title{Temperature dependent piezo- and pyroelectric coefficients in \zro\ from machine learned force fields}
\title{Piezo- and pyroelectricity in Zirconia: a study with machine learned force fields}

\author{Richard Ganser}%
\affiliation{Department of Applied Sciences and Mechatronics, Munich University of Applied Sciences, Lothstr. 34, 80335 Munich, Germany}%
\affiliation{These two authors contributed equally to this work.}

\author{Simon Bongarz}
%\email{robin.materlik@hm.edu}
\affiliation{Department of Applied Sciences and Mechatronics, Munich University of Applied Sciences, Lothstr. 34, 80335 Munich, Germany}%
\affiliation{These two authors contributed equally to this work.}

\author{Alexander von Mach}%
\affiliation{Department of Applied Sciences and Mechatronics, Munich University of Applied Sciences, Lothstr. 34, 80335 Munich, Germany}%

\author{Luis Azevedo Antunes}%
\affiliation{Department of Applied Sciences and Mechatronics, Munich University of Applied Sciences, Lothstr. 34, 80335 Munich, Germany}%

\author{Alfred Kersch}
\email{alfred.kersch@hm.edu}
\affiliation{Department of Applied Sciences and Mechatronics, Munich University of Applied Sciences, Lothstr. 34, 80335 Munich, Germany}%

\date{\today}% It is always \today, today,
             %  but any date may be explicitly specified

\begin{abstract}
The discovery of very large piezo- and pyroelectric effects in \zro and \hfo-based thin films opens up new opportunities to develop silicon-compatible sensor and actor devices. The effects are amplified close to the polar-orthorhombic to tetragonal phase transition temperature.
Molecular dynamics is the preferred technique to simulate such effects, though its application has to solve the dilemma between sufficient accuracy and sufficient efficiency of the interatomic force field.
Here we present a deep neural network-based interatomic force field of \zro\ learned from ab initio data using a systematic learning procedure in the Deep Potential framework. The model potential is verified to predict a variety of structural and dynamic properties with an accuracy comparable to density functional theory calculations. 
Then the Deep Potential model is used to reproduce the different thermal expansion and piezo and pyroelectric phenomena in \zro\ with molecular dynamics calculations. At low temperature simulating the direct effect we find negative values for the piezo-and pyroelectric coefficients matching the {\it ab initio} calculations. Approaching the phase transition temperature these values remain negative and become large. Simulating the field induced effect above the phase transition temperature we find positive, giant piezo-electric coefficients matching the observations. The model is able to explain the large values and the sign of the experimental observations in relation to the polar-orthorhombic to tetragonal phase transition. The model furthermore explains the recently observed giant dielectric constant in a similiar system.

%The validity of the model potential is discussed and the systematic improvement in accuracy is suggested.
\end{abstract}

%\pacs{Valid PACS appear here}% PACS, the Physics and Astronomy
                             % Classification Scheme.
%Use showkeys class option if keyword
                              %display desired
\maketitle

\section{\label{sec:Introduction}Introduction}

%Introduce Abbreviations DP (Deepmodel) and DFT. 

Ferroelectrics based on \hfo\ and \zro\cite{Book2019} are attracting increasing attention due to their great potential for a variety of applications such as next-generation memories\cite{miko}, nanotransistors\cite{hoffmann}, and piezoelectric and pyroelectric\cite{kirbach-piezo,jachalke} thin-film devices. The ferroelectric phase\cite{naturereview} has been recognized as the polar-orthorhombic $Pca2_1$, no 29 (po-phase), which does not stabilize in this material system without special conditions, but requires very thin layers\cite{Hoffmann2015} or specific doping\cite{Batra2017,Materlik2018}. In this way, both \hfo\ and \zro\ can act as a base material. 
While ferroelectric devices depend on the existence of a stable and and switchable polarization, the piezoelectric and pyroelectric properties result from a sufficient electrotromechanical and electrothermal response of the polarization. Experiments showed that piezoelectric coefficients obtained with dopant optimized material are attractively large compared to other thin film materials. Interestingly, large positive values up to $d_{33}$ = 73pm/V were reported\cite{kirbach-piezo} as well as large negative values\cite{mart-piezo} of $d_{31}$ = -11pm/V.
Very large values could be theoretically understood by the electro-strain effect\cite{Jo2012, Hao2019}, which explains the response as a field-induced phase transition from the
tetragonal $P4_2/nmc$, no 137 (t-phase) and the associated volume change\cite{falkowski}, but the different signs are puzzling. The large values furthermore appear to mask the genuine piezoelectric coefficient, which is anomalously negative but small, as indicated by ab initio calculations\cite{liu, dutta}.

The amplified response of the material near the phase transition has also been observed in the pyroelectric effect. The giant pyroelectric response up to $\Pi_3=-1300\mu C m^{-2}K$ was found in polycristalline, doped thin films stimulated by an electric field at different temperatures\cite{Hoffmann-pyro}. Temperatures were chosen near the phase transition temperature between the ferroelectric and paraelectric phases, whose Curie temperature could be varied with doping concentration. The giant effect was smeared over a range of more than 100K, and for explanation it was suggested that the Curie temperature also depends on an effect of grain size on the free energy\cite{Materlik2015, kuenneth}, which creates a window of enhanced values and makes the effect more useful for applications.
A very large pyroelectric response was also found in similiar samples with low-frequency thermal stimulation experiments without an electric field\cite{mart}, which resembles the typical pyroelectric application. In these experiments, an influence of the phase transition on the pyroelectric effect is not directly apparent, but was suggested.
Liu et al.\cite{liu} studied the pyroelectric effect for \hfo\ with ab initio harmonic lattice theory developed in \cite{pantelides} and for \hfo\ and Si:\hfo\ with ab initio molecular dynamics (AIMD). They found that the primary $\Pi_1$ coefficient (constant external strain) and the secondary $\Pi_2$ coefficient (strain due to thermal expansion) both contribute with a negative sign and therefore add up to the large effect. The contribution of the secondary coefficient is explained by the anomalous negative genuine piezo effect. The contribution of the primary coefficient is related to anharmonic effects of the lattice, although this is only beginning to be understood in the harmonic approximation. To better investigate the anharmonic effects on the primary coefficient, Liu et al. performed ab initio molecular dynamics calculations, which are computationally expensive. They found that the enhancement of the secondary coefficient is associated with the ferroelectric to paraelectric phase transition.

This finding is the starting point for our investigation. The AIMD is the optimal tool to study the pyroelectric effect for anharmonic lattice effects. However, progress is hindered by large computation times. We have therefore developed a machine-learned deep potential (DP) for \zro, which has been missing in the literature, and which combines near ab initio accuracy with significantly lower computational time. We were motivated to do this by the development of a DP for \hfo\ by Wu et al.\cite{wu}. Using the developed DP for \zro, we will calculate the piezoelectric and pyroelectric coefficients for \zro\ including the temperature dependence. We will evaluate the results for very low temperatures, intermediate temperatures, and temperatures near the transition from the ferroelectric to the paraelectric phase.

\section{\label{sec:Methods}Methodology}

%\subsection{\label{sec:Computational}Computational Methods}

%QUANTUM ESPRESSO with PBE-SOL, DEEPMD-KIT, LAMMPS, AIMS
%What tool is missing?
The Deep Potential (DP) method for fitting the potential energy surface has been developed, validated and described in the literature\cite{Han, Zhang}, so in this section we only briefly introduce the idea, the generation of the database from first-principles calculations and the training and simulation protocols used in this work. The DP decomposes the total energy of a system into atomic contributions $E = \sum_i E_i$ where $E_i$ depends only on atom $i$ and the $m$ neighbours $j$ when the relative position $r_{ij}=r_j-r_i$ is smaller than a cutoff radius $r_c$. DP defines the energy contribution by $E_i=\mathcal{N}^{\omega}(\mathcal{D}_{i}(\mathcal{R}_i))$, where $\mathcal{R}_i=(r^T_{ij_1},r^T_{ij_2},...,r^T_{ij_m})^T$ is the environment matrix whose entries are embedded into the descriptor matrix $\mathcal{D}_i$ containing symmetry preserving expressions of the coordinates. The embedding net $\mathcal{R}_i$ and the fitting net $\mathcal{N}^{\omega}$ have trainable parameters $\omega$ which maps the descriptor to $E_i$. $\mathcal{D}_i$ and 
$\mathcal{N}^{\omega}$ are smooth functions which allows to calculate the atomic force as negative gradient of the total energy. In our model we use a cut-off radius of $6 \AA$ and and the predefined neural network size.

In the training of the model the discrepancy between the prediction of DP and the total energy and atomic forces calculated by an {\it ab initio} approach is minimized with respect to the model parameters $\omega$
\begin{equation}
    \underset{\omega} {min}  {p_e |E-E^{\ast}|^2 + \frac{p_f}{3N} |F_i-F_i^{\ast}|^2 } 
\end{equation}
where $E^{\ast}$ and $F_i^{\ast}$ denote the {\it ab initio} total energy and atomic forces, respectively. $p_e$ and $p_f$ are tunable prefactors where $p_e$ progressively increases and $p_f$ decreases.

\subsection{\label{sec:Dataset}Training Data}

To train the DP model, a large training data set must be created.
For the training data set the monoclinic m-phase ($P2_1/c$, no 14), polar-orthorhombic po-phase ($Pca2_1$, no 29) and tetragonal t-phase ($P4_2/nmc$, no 137) in a 12-atomic pseudo-cubic configuration were used. 
To create a first data set which represents vibrational motion with an average energy related to a specified temperature, we displaced the ionic coordinates about the equilibrium position according to the following rule. The potential energy for each 
displacement $\delta u_{\mu \alpha}$ from the equilibrium position $u_{\mu \alpha}$ was estimated, with $\mu$ the atomic label and $\alpha$ the direction x, y or z. The estimate for a maximal displacement uses the Interatomic Force Constants $\Phi_{\mu\alpha,\nu\beta}$, which have been calculated from {\it ab initio} for each crystal phase, as
\begin{equation}
    4 k_B T = E_{max} \geq E_{\mu\alpha} = \Phi_{\mu\alpha,\mu\alpha} (\delta u_{\mu \alpha}^{max})^2
    \label{threshold}
\end{equation}
The maximum allowed energy per degree of freedom is larger than the expected average by the empirically chosen factor 8, which allows strongly deformed structures for the training data set, but on the other hand represents a cutoff against too strongly deformed structures.
%The maximal allowed energy per degree of freedom is by a factor 8 larger than the expected average, which was found to be an efficient choice.
The equation can be resolved to $\delta u_{\mu \alpha}^{max}$ for each degree of freedom. The overall covered energy range of the data sample is first defined with a randomly, uniform choosen temperature $T$ smaller than a threshold temperature $T_{max} < 2500K$ in the order of the melting temperature. The actual displacement $\delta u_{\mu \alpha}$ for each degree of freedom was then randomly chosen to be in the range $[-\delta u_{\mu \alpha}^{max},+\delta u_{\mu \alpha}^{max}]$.
In this way, ionic positions which may occur in a vibrational motion are homogeneously distributed over the configuration space in a way which takes care of soft mode oscillations, performs an energy cutoff, and because of the quadratic dependence in (\ref{threshold}) generally favours low energy displacements.
The drawback of this method is that the data do not contain any information about cell  deformations. 
Therefore a second data set was created which includes cell deformations. For this purpose strain tensor data $\epsilon_{ij}$ were created to calculate new lattice vectors $\mathbf{\tilde{a}_i}$ from the equilibrium lattice vectors $\mathbf{a_j}$ from
$$\mathbf{\tilde{a}_i}=(\delta_{ij}+\epsilon_{ij})\mathbf{a_j}$$
The values of $\epsilon_{ij}$ were uniformly sampled in a range of $\pm 5 \: \mathrm{\%}$ for all elements. With these values compression and stretching of the cell are well covered, as well as shear deformation which occurs during the phase transition from no 14 to no 137 or no 29.

A third data set combines the atomic and the lattice perturbation. A smaller lattice perturbation (0.5\% for shear and 3\% for linear strain) was combined with atomic perturbation. It was found that the third data set improves results for the elastic constants.

In total 90 000 structures were included in the data set with 30000 structures for each of the three phases and 10000 for each of the three subsets.

In the comparable model development of Wu et al.\cite{wu} 21768 structures of large 2x2x2 super cells (96-atoms) were included into the training process. Larger cells contain more relevant information, but are also significantly more computationally intensive, as the computing time grows at least linearly with the volume. In this paper we investigate a model based on smaller structures which can be improved subsequently with larger structures containing long range correlations or with more irregular structures containing surfaces and defect configurations.

For all structures of our data set the total energy was calculated from {\it ab initio} with quantum-espresso\cite{espresso} (QE) using the PBEsol\cite{pbesol} exchange-correlation functional. We choose efficient ultrasoft poseudopotentials with an energy cutoff of 450eV and a 4x4x4 k-point sampling. The results were compared with a highly accurate all electron implementation of the PBEsol functional in FHI-aims with a 6x6x6 k-point sampling, with an agreement of about 1meV/f.u., see Table \ref{tab1}.

\subsection{\label{sec:Training} Training}
For the training process we used the DeepMD-kit \cite{Wang2018} with an implementation of the Deep Potential (DP).
We used the standard settings of the most important parameters: a batchsize = 2 000 000, sizebatch = 50 000, learning decay rate = 0.95 , learning steps = 10 000 and start learning rate = 0.005. The three data sets were learned subsequently: after learning the first data set, the result was used as initialization for the learning on the second data set. Similiarly for the third data set.
\begin{figure}[h!]
\centering
\includegraphics[width=9.0cm]{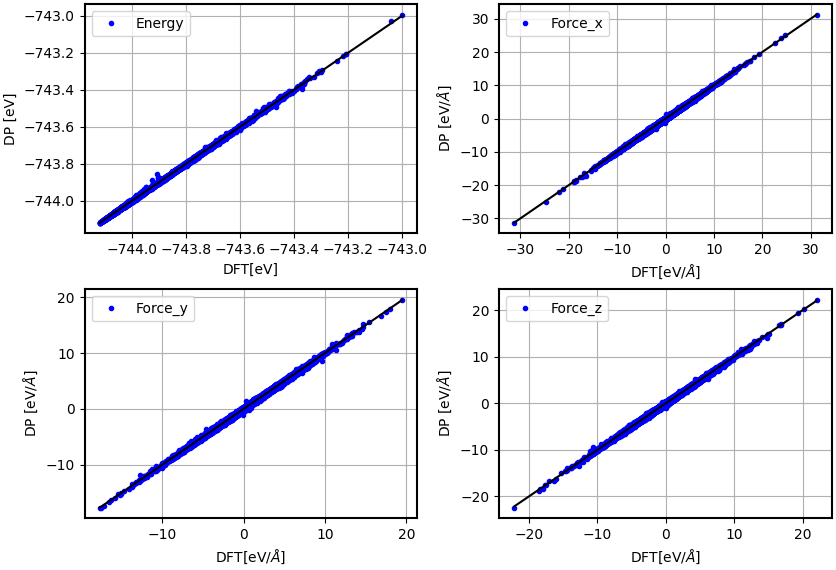}
\caption{Test of the model after the final training stage: The total energies and atomic forces predicted by DP in comparison with the DFT results from QE.}
\label{fig:Training}
\end{figure}

Fig. \ref{fig:Training}  shows the fit of the energy to the training data.
In the molecular dynamics simulation of the paper the model is used within the range of the training data. In the validation the range outside the training data is also explored to show the limitation of the model.

\subsection{\label{sec:Validation}Validation}

To validate the \zro\ DP model the DeepMD-kit \cite{Wang2018} was used to predict the total energy of various crystalline structures. Table \ref{tab1} compares the lattice parameter of different phases of \zro\ optimized with DP and DFT, demonstrating the excellent agreement. The table contains the comparison of the total energy of relaxed structures between DP and DFT as well as total energies obtained from highly accurate calculation with the all-electron code FHI-aims\cite{aims}. The data include the three phases included in the learning process, the second ferroelectric phase $Pmn2_1$ no 31, as well as three 24-atomic phases: the two inequivalent nonpolar  $Pbca$ no 61 and antipolar orthorhombic $Pbca$ no 61x\cite{pssground}.

\begin{table}[]
 \caption{Lattice parameters (a,b,c) and lattice angle $\beta$ at 0 K calculated with DP (bold letters) and DFT. The calculated energies are relative to the monoclinic case and calculated with DP (bold), QE as well as FHI-aims using PBEsol.}
\begin{tabular}{|l|l|l|l|l|l|l|}
\hline
                & a                & b               & c               & $\beta$         & $\Delta$E(QE)       & $\Delta $E(aims) \\
                & [\AA]            & [\AA]           & [\AA]           & [$^{\circ}$]    & $\Delta$E$\mathbf{(DP)}$       &  \\ \hline
$P2_1/c$        & 5.1              & 5.22            & 5.22            & 80.3            & 0.0              & 0.0              \\ 
no14  m-phase & $\mathbf{5.01}$  & $\mathbf{5.21}$ & $\mathbf{5.22}$ & $\mathbf{80.4}$ & $\mathbf{0.0}$   &                  \\ \hline
$Pbca$ nonpolar & 10.17            & 5.19            & 5.30            & 90              & 24.9             & 25.3             \\ 
no 61   & $\mathbf{10.16}$ & $\mathbf{5.19}$ & $\mathbf{5.29}$ & $\mathbf{90}$   & $\mathbf{20.6}$  &                  \\ \hline
$Pbca$ antipolar  & 10.05            & 5.25            & 5.25            & 90              & 42.2             & 41.7             \\
no 61x  & $\mathbf{10.01}$ & $\mathbf{5.25}$ & $\mathbf{5.07}$ & $\mathbf{90}$   & $\mathbf{81.7}$  &                  \\ \hline
$Pca2_1$        & 5.05             & 5.26            & 5.07            & 90              & 53.4             & 52.9             \\ 
no 29  po-phase         & $\mathbf{5.05}$  & $\mathbf{5.26}$ & $\mathbf{5.07}$ & $\mathbf{90}$   & $\mathbf{58.8}$  &                  \\ \hline
$P4_2/nmc$      & 5.07             & 5.18            & 5.07            & 90              & 80.4             & 78.5             \\ 
no 137  t-phase        & $\mathbf{5.07}$  & $\mathbf{5.18}$ & $\mathbf{5.07}$ & $\mathbf{90}$   & $\mathbf{81.7}$  &                  \\ \hline
$Pmn2_1$        & 3.46             & 5.18            & 3.75            & 90              & 95.3             & 94.2             \\ 
no 31           & $\mathbf{3.47}$  & $\mathbf{5.18}$ & $\mathbf{3.75}$ & $\mathbf{90}$   & $\mathbf{109.7}$ &                  \\ \hline
\end{tabular}
\label{tab1}
\end{table}

\begin{figure}[h!]
\centering
\includegraphics[width=9.0cm]{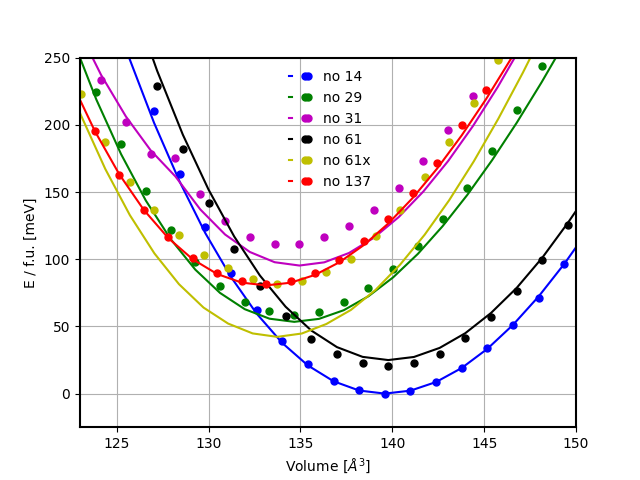}
\caption{Energy-volume curves with DP (dots) and DFT (lines) for the phases from Table \ref{tab1}.}
\label{fig:EV}
\end{figure}

The next validation is done with energy-volume curves. The unit cells of Table \ref{tab1} were strained and the ionic positions were subsequently relaxed in DFT. Then the energy of the structures was calculated with DFT and DP. Fig. \ref{fig:EV} shows the results.

The comparison is excellent for the phases involved in the learning process and also for the no 61 nonpolar, fair for the polar phases no 31 and no 7, and the antipolar no 61 with an energy difference of about 20meV/f.u. The DP model is capable to predict metastable crystal phases not participating in the learning process, but with an error in energy. This error seems to be related to long range dipole interaction contained in the antipolar no 61 structure, and to a lesser extent in the polar no 7 and no 31 structure.

The next validation of the DP model is the strain tensor and compliance tensor, which is contained in the energy-volume curve. The tensor furthermore relates the piezoelectric stress and strain coefficients. The values from DP and DFT compare excellent and are found in the Supplemental Material S1.

\begin{figure}[h!]
\centering
\includegraphics[width=9.0cm]{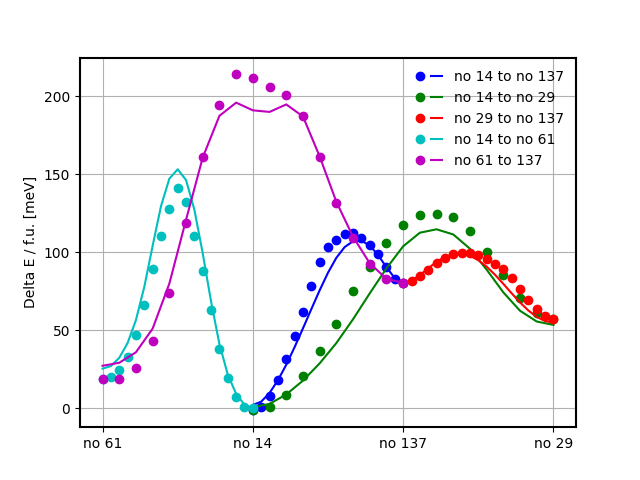}
\caption{Minimal energy paths between phases no 61, no 14, no 137 and no 29. The lines are representing the DFT results and the dots the energy of the frames calculated with the DP model.}
\label{fig:NEB}
\end{figure}

A further validation concerns the minimum energy curves that connect the discussed phases. We have calculated the minimal energy paths with the Nudged Elastic Band (NEB) method with FHI-aims and the PBEsol functional. The energy of the 19 frames connecting the crystal phases were calculated with the DP model. Fig. \ref{fig:NEB} shows the comparison of the data. The crystal structures from the minimal energy paths extend to the energy landscape far away from equilibrium structures. The comparison is excellent.

\subsection{\label{sec:Molecular Dynamics}Molecular Dynamics}

The DeePMD-kit is interfaced with the high-performance classical molecular dynamics packages LAMMPS\cite{Plimpton1995}. Thus the DP potential energy models can be used to perform efficient MD simulations for different purposes. 

The Born-Oppenheimer approximation is a sufficient approximation for MD of crystalline properties at elevated temperatures but fails to capture low temperature effects related to zero-point vibrations. In order to take zero point vibration and quantum occupation into account, a quantum-thermal bath (QTB) is coupled to the MD simulations as proposed by Dammak\cite{Dammak2009}. By adding a random and a dissipative force term, which follows the power spectral density given by the quantum fluctuation-dissipation theorem, a model of the Debye-behaviour close to 0K can be achieved. The quantum-thermal bath needs the Debye frequencies of the associated crystal structures, which were calculated using the phonopy-package, see below, and were obtained as 9.345THz for the no 29 po-phase, 8.119THz for the no 137 t-phase and 8.52THz for the no 14 m-phase. The parameters for the thermostat and barostat are choosen to be uniform in all simulations with $p_{damp}$ being 1ps and $t_{damp}$ being 100fs. 
The values are chosen to be relatively large to prevent overshooting of the atomic deflections in the simulation, which could potentially lead to phase transitions and other spurious effects.
The supercells are composed of 4x4x4 12 atomic unit cells, corresponding to 768 atoms. The initial supercells are fully relaxed, using the conjugate gradient (CG) algorithm of LAMMPS, which allows the relaxation the crystal cell, until the forces converge below 10$^{-4}$eV/\AA\  and energies below 10$^{-6}$eV.
%\begin{comment}
%The equations of motions are the ones proposed by Shinoda.\cite{Shinoda2004} with the strain energy as proposed by %Parrinello\cite{Parrinello1998}.
%\end{comment}

The MD simulations are conducted with a time step of 1fs at constant temperatures ranging from 1K to 1200K, in 100K intervals. 
The microcanonical ensemble (NVT) simulation cells are constructed by first using isothermic-isobaric ensemble (NPT) simulations at different temperatures and zero stress to determine the thermal expansion coefficients relative to the relaxed cell \cite{Klarbring2018}. By enlarging the relaxed structure corresponding to the desired temperature and thermal expansion coefficients, a shortening of the equilibrium times was achieved.
The simulations were conducted with a thermal equilibrium time of 5ps and a sampling time of at least 20ps.
The polarization is calculated by:
\begin{equation}
    P = \frac{\delta}{2\Omega_0}N\,Z^{\ast}
    \label{eq:polarization}
\end{equation}
$\delta$ being the displacement between the zirconium and the oxygen ions centers of mass, N the number of atoms in the unitcell, $Z^*$ the Born charges, and $\Omega_0$ the size of the supercell. The Born charges were calculated using ABINIT with the PBEsol functional, which resulted in averaged diagonal values of q = 5.10e for zirconium ions and q = -2.55e for oxygen ions.
The interaction with an uniform, external electric field is calculated from multiplication with the Born charges.\\
As an error estimate, values over 100 fs were averaged to one data-point. These were then treated as stochastic independent results, which allows for gaussian error estimation.

\subsection{\label{sec:Phonon dispersion}Phonon dispersion}

The phonopy package\cite{Togo2015} was used for the calculation of the phonon-dispersion relation, for validation of the energy landscape around the minima of metastable phases, and for the calculation of the free energies using the harmonic approximation. To achieve sufficient accuracy the structures were relaxed until the forces and energies converged below 1e-5eV/\AA\ and 1e-6eV. This was conducted with the DFT implemented in FHI-aims with the PBEsol functional and with the DP model. The phonon dispersion of the no 14 m-phase, no 29 po-phase, and no 137 t-phase were calculated using supercells of size 12.  DFT and DP compare very well, which is documented in the Supplemental Material S2.

In total we have used 3 different models: DFT calculations, DP model calculations with the potential learned from DFT, and MD calculations using the DP model.

\section{\label{sec:New_Results}Results}

\subsection{Curie-temperature}

Using the phonon-dispersions, the temperature-dependent total energy $E(T)$ and entropy $S(T)$ were calculated relative to the m-phase, which provides the free energy $F(T) = E(T) - T\,S(T)$. The intersection of the free energies $F(po)$ and $F(t)$ defines the value of the Curie-temperature $T_C$.

\begin{figure}[h!]
\centering
\includegraphics[width=9.0cm]{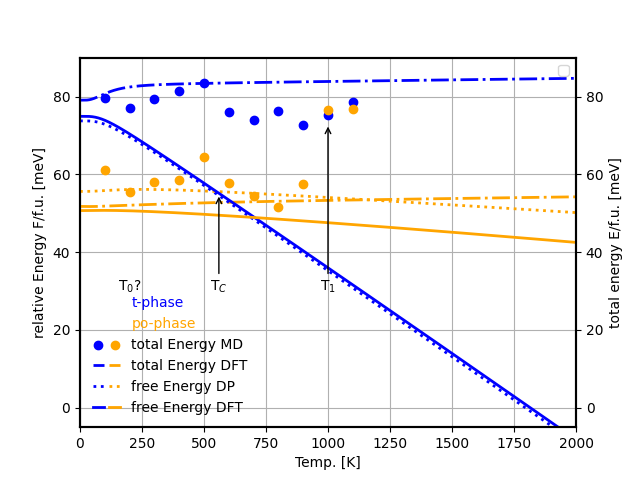}
\caption{The blue lines and symbols display the values of the t-phase and the orange lines and dots the values of the po-phase. All values are relative to the calculated energies of the m-phase in DP and DFT.}
\label{fig:free}
\end{figure}

Fig. \ref{fig:free} shows the comparison of the free energies $F(T)$ relative to the m-phase for DFT and DP calculations. The data compare well, but the DP energy of the  po-phase is about 5 meV/f.u. higher than the ab initio calculation indicating a training error. This error shifts the Curie-temperature $T_C$ to lower value. For the DP model we estimate $T_C$ = 560K, whereas the reported ab initio values are around $T_C$ = 650K\cite{Azevedo2022}.

Because the computation of $S(T)$ with the MD method is intricate, we calculated only the total energy $E(T)$ with MD and compared the resulting statistical averages with values from phonopy using DFT forces.
Fig. \ref{fig:free} shows the comparison of the total energies E(po) and E(t) relative to the m-phase from DFT calculations and from MD with DP. The MD results compare well to the DFT results within the gaussian error.
Remarkably, the phase transition appears in $E(T)$ as a discontinuity at the phase transition temperature $T_1$. Interesting is that this temperature with $T_1$ = 1000K is larger than the Curie-temperature. Such a behaviour can be expected from the energy landscape when in addition to the lowering of the t-phase free energy with temperature there is an energy barrier present similiar to Fig. \ref{fig:NEB} which persists at elevated temperatures. A consequence is a bidirectional thermal hysteresis for the phase transitions between the po- to t-phase. A thermal hysteresis in \hzo\ for different stoichiometries of 50-200K has recently been found experimentally\cite{Schroeder2022}. Such a thermal hysteresis is expected for ferroelectric materials and contained in a 6th order Landau-Devonshire effective model \cite{Strukov}. But a phase transition temperature $T_0 < T_C$ for the reverse transition from the t- to the po-phase is also expected from the experimental observation, but not visible in the MD data even at longer simulation times of 50ps, although the ferroelectric phase has a lower free energy as prerequisite. The difficulty to describe the formation of the po-phase in spite of a missing ferroelectric instability, which would be visible in the phonon dispersion, has been discussed by Reyes-Lillo et al.\cite{ReyesLillo2014} and recently Delodovici\cite{Delodovici}. The authors identified a multi-phonon coupling as a possible explanation for the transition to the po-phase. Therefore this transition could require a much longer simulation time.

At this point the values of the free energies should be discussed in relation to experimental data. Crystalline \zro\ is mostly found either in the  m- or in the t-phase. The t-phase is substantially favoured in polycrystaline nanoscale thin films. Garvie\cite{Garvie1965} proposed this to be an effect of a lower surface energy of the t- relative to the m-phase. This observation has been extended\cite{Materlik2015, kuenneth} to the po-phase with a prediction of stability windows for the o-phase and the t-phase, depending on the size of the nuclei, which is experimentally confirmed\cite{mimura}.
Furthermore, \zro\ at room temperature is t-phase and the po-phase can be stabilized with reduced temperature \cite{nanoletters, cheng}.

The conclusion for our paper is that the calculated free energy is a model for mono-crystalline or
large-grain \zro\ with negligible surface effects, and that the free energy in polycrystalline material is shifted by surface energy effects, resulting in t- or po-phase \zro\ at room temperature, and a Curie temperature around room temperature.

Polycrystalline, Si-doped \hfo\ with about 5\% doping has similarities to polycrystalline \zro\ regarding the free energy because the t- and po-phase are competing at room temperature\cite{Hoffmann-pyro, richter}. In both materials the Curie-temperature is close to room temperature, whereas in our DP \zro\ the Curie-temperature is shifted to about 560K.

After the preliminary investigation of the energy landscape, we conclude that the trained DP model compares well with the DFT results. We have therefore enabled efficient MD simulation for the potentially ferroelectric \zro.

\subsection{\label{sec:giant}Field induced phase-transition and giant piezoelectric effect}

The antiferroelectric behaviour in \zro\cite{Mueller2012c} has been interpreted as field induced phase transition. The prerequisite is that the free energy of the t-phase is below the ferroelectric phase. While this should be true for the poly-crystalline \zro, according to ab initio models this is not true for the mono-crystalline material. Furthermore, the t-phase shows no instability from thermal motion within simulation time. With MD we demonstrate that an applied electric field forces an instability and a subsequent phase transition. The electric field values in Table \ref{tab2} were increased in 0.1MV/cm steps until the phase-transition occurred for each temperature. The size of the field ${\cal{E}}_1$, for the t- to po-phase transition, is nearly constant and is consistent with the size of the energy barrier of about 20meV/f.u. (see Fig. \ref{fig:NEB}) due to the field energy $W_1=\Omega{\cal{E}}_1 P_{TS}$. $\Omega$ is the f.u.-volume and $P_{TS}$ the polarization at the transition state, derived from NEB calculation to be about $0.4P$. Interestingly, the required field-strength for the phase transition remains constant for temperatures above $T_1$,  where the free energy of the t-phase lies below the po-phase, and we observe in fact a field induced phase transition.
The experimentally measured electric field strengths required for phase transition are in the range of 1-2MV/cm compared to the 3-4MV/cm from our simulation. Although these values are in qualitative agreement, the reasons for the remaining discrepancy are of interest. Possible explanations include the choice of the PBEsol functional compared to others with different energy barriers (see e.g. Guan et al.\cite{Guan}), a possible effect of surface energy of a polycrystal on the energy landscape, and lowered energy barriers from intrinsic defects.

The field induced phase transition followed by a volumetric change is the explanation for the giant piezoelectric effect, also proposed to explain the values in many high-strain piezoelectric ceramics\cite{Jo2012, Hao2019}.
Beyond $T_1$ the field induced transition to the po-phase reverses to the t-phase when the electric field is reduced, as shown in Fig. \ref{fig:free}.
Falkowski et al.\cite{falkowski} have investigated the volumetric change for doped \hfo\ and \zro\ and estimated piezo-strain coefficients assuming a required field of 2MV/cm, and obtained a value of 29pm/V in very good agreement with data\cite{Starschich2017}. In Table \ref{tab2} we have added the volumetric change in our MD calculation. The resulting piezo-strain coefficient is positive and has a value beyond $T_1$ of 10 pm/V, which is in good agreement with Falkowski taking our larger electric field from the simulation into account.

\begin{table}[]
\caption{Applied electric field for the phase transition from t-phase to po-phase and the associated volume increase. Furthermore the field for the phase transition from po-phase to t-phase.}
\begin{tabular}{|c|c|c||c|}
\hline
\begin{tabular}[c]{@{}c@{}}Temp\\ {[}K{]}\end{tabular} & \begin{tabular}[c]{@{}c@{}}${\cal{E}}_1$-field\\  no137 -\textgreater no29\\  {[}MV/cm{]}\end{tabular} & \begin{tabular}[c]{@{}c@{}}\\volume \\  change \\ {[}\%{]}\end{tabular} & \begin{tabular}[c]{@{}c@{}}${\cal{E}}_2$-field\\  no29 -\textgreater no137\\ {[}MV/cm{]}\end{tabular} \\ \hline
300                                                    & 3.5                                                                                            & 0.54                                                                    & 4.7                                                                                           \\ \hline
400                                                    & 3.6                                                                                            & 0.47                                                                    & 4.2                                                                                           \\ \hline
500                                                    & 3.6                                                                                            & 0.41                                                                    & 3.2                                                                                           \\ \hline
600                                                    & 3.5                                                                                            & 0.36                                                                    & 2.9                                                                                           \\ \hline
700                                                    & 2.9                                                                                            & 0.32                                                                    & 2.3                                                                                           \\ \hline
800                                                    & 3.2                                                                                            & 0.30                                                                    & 1.5                                                                                           \\ \hline
900                                                    & 3.3                                                                                            & 0.28                                                                    & 0.9                                                                                           \\ \hline
1000                                                   & 3.1                                                                                            & 0.28                                                                    & $\sim$0                                                                                       \\ \hline
1100                                                   & 3.3                                                                                            & 0.29                                                                    & 0                                                                                             \\ \hline
1200                                                   & 3                                                                                              & 0.31                                                                    & 0                                                                                             \\ \hline
\end{tabular}
\label{tab2}
\end{table}

Whereas the giant piezo-strain effect has the t-phase as initial condition, the conventional piezoelectric effect has the po-phase as initial condition, which leads to negative coefficients. To study the energy landscape, it is of interest to apply a strong external field ${\cal{E}}_2$ opposite to the polarization direction causing a phase transition to the t-phase. For low temperatures, the free energy of the po-phase is below the t-phase, and the electric field energy has to overcome the free energy difference plus the height of a possible energy barrier. Table \ref{tab2} shows the required fields, linearly decreasing with temperature from  4MV/cm at 300K to 0MV/cm at $T_1$.
The field energies to transition the po- to the t-phase compare well to a decrease of the free energy with increasing temperature. The field energy $W_2=\Omega{\cal{E}}_2 P_{po}$ fits well to the barrier height of 50meV/f.u. for the path from po-phase to t-phase (see Fig. \ref{fig:NEB}).
In contrast to the t-phase, the po-phase shows an intrinsic instability when the phase transition temperature $T_1$ is approached. This is crucial for a deeper understanding of the piezo-, pyro-, and dielectric-effects close to the phase transition temperature. Although our model does describe a mono-crystalline instead of a poly-crystalline \zro, we think that the surface effects mainly cause a shift of the po- to t-phase transition temperature\cite{Materlik2015}.

%Between 400 K and 1000K the no 29 phase can be triggered to decay into the no 137 phase by applying an electric field and above 1000 K the no 29 can be stabilized starting from the no 137. \\
%The presented table shows the required electric field strength to trigger the phase transition from the no 29 to the no 137 and from the no 29 to the opposite polarized no 29. Between 400 K and 1000 K the re-polarization happens in two steps, first the decay from the no 29 to the no 137, staying at this plateau and then transitioning to the opposite polarized no 29. Below 400 K the repolarization path follows the same pattern, but doesn' t halt in the no 137 structure. \\
%Due to the different cell sizes the phase-transition leads to a volume reduction, resulting in a electric field direction dependent gigantic piezo-eletric coefficient. \\
%Above 1000 K the no 137 phase, by applying an electric field to the sample, transforms into the no 29, which is known as the anti-ferroelectric behaviour. The direction of the electric field has no influence on the piezoelectric effect, only changing the polarization direction, yielding an, electric field direction independent gigantic piezoelectric coefficient.
%This effect can not be completely reproduced by straining the cells. Exposing the no 29 to a positive strain, resulting in an enlargement, triggers the decay into the no 137, which is contra-intuitive due to the smaller size of the no 137. Straining the tetragonal cell produces, in the range up to $\pm 1\%$ strain, no instability. 

\subsection{\label{sec: Thermal expansion of polar-orthorombic phase}Thermal expansion of polar-orthorhombic phase}
\begin{figure}[h!]
\centering
\includegraphics[width=9.0cm]{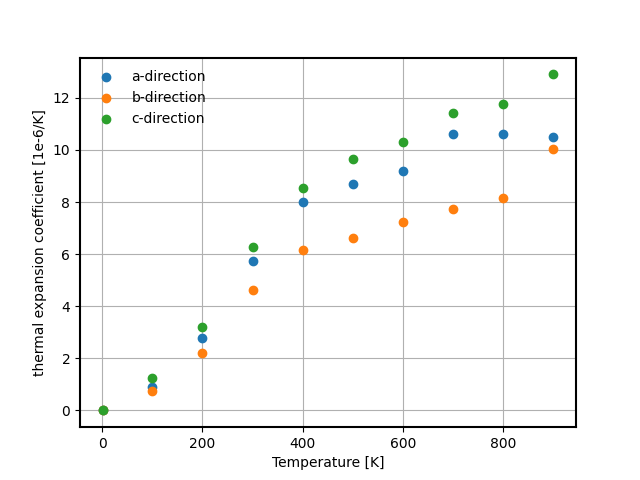}
\caption{Anisotropic, temperature dependent thermal expansion coefficient of po-phase \zro\ from MD.}
\label{fig:expansion}
\end{figure}

To explore the influence of temperature on the crystal volume, we calculated the thermal expansion coefficients for the t-phase and the m- phase using NPT ensembles, which are in good agreement with experimental data\cite{Haggerty2014}, see Supplemental Material S3. Based on the agreement of these simulations with the experiments, we calculated the anisotropic thermal expansion coefficients for the po-phase up to $T_1$ plotted in Fig. \ref{fig:expansion}. It is important to note that similar to the heat capacity, the coefficients of thermal expansion of crystalline materials must generally vanish when approaching the temperature zero point. Our classical MD calculations reproduce this behaviour only because we use the QTB model, and therefore our predictions match the data 
well also at low temperatures.

\subsection{\label{sec: Piezoelectric stress coefficient}Piezoelectric stress coefficients}
\begin{figure}[h!]
\centering
\includegraphics[width=9.0cm]{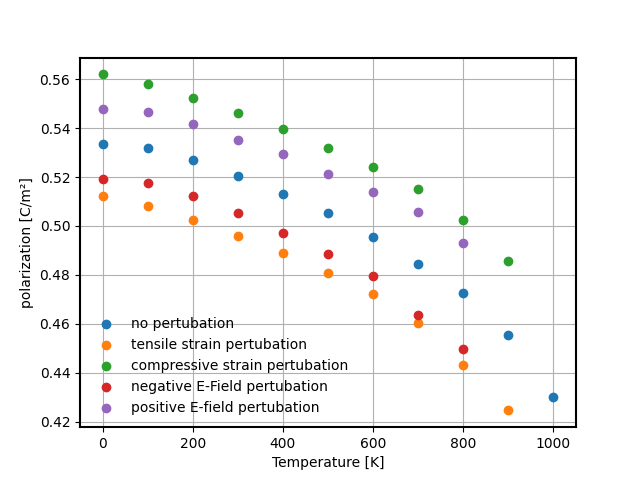}
\caption{Polarization of the po-phase against temperature, exposed to $\pm 1MV/cm$ electric fields and $\pm 1\%$ strain perturbation in polarization direction.}
\label{fig:polarization}
\end{figure}

We performed NVT simulations with up to $\pm1\%$ applied uniaxial strains and electric fields up to $\pm$1MV/cm along the polarization direction at various temperatures, and determined the polarization.
The results in Fig. \ref{fig:polarization} show generally a decrease of polarization with temperature, and a dependence on strain and electric field perturbation.
From such results the pyroelectric, piezoelectric and dielectric coefficients are derived. The polarization of the unperturbed po-phase was calculated in MD simulations with Eq. (\ref{eq:polarization}) to be 0.69C/m$^{2}$, which is larger than the values calculated with the Berry phase method\cite{ReyesLillo2014}. This is because of the in (\ref{eq:polarization}) inherent simplified assumption of constant Born charges along the path to the paraelectric reference phase. For a better comparison of our data with experiments and ab initio calculations, we scaled our polarization values to 0.545C/m$^{2}$, which coincides with the Berry phase calculation with PBEsol from ABINIT.

The piezoelectric stress coefficients\cite{Wu2005} describe the change of polarization component $P_{\alpha}$ with applied strain $\eta_j$ ($j$ Voigt index) at zero electric field $\cal{E}$
\begin{equation}
    e_{\alpha j}= \left.\frac{\partial P_\alpha}{\partial \eta_j}\right\vert_{\cal{E}}
\end{equation} 

In {\it ab initio} calculations \hfo\cite{liu, dutta} and \zro\ show negative piezoelectric coefficients (see Table \ref{tab3} and Supplemental Material S1), opposed to positive relations between polarization and strain typically observed in ferroelectric perovskites.
Dutta\cite{dutta} et al. explained the negative sign in \hfo\ from the peculiar chemical bond of the polarizing oxygen atoms to the neighboring Hf atoms, which seems to be similar in \zro. 

\begin{table}[]
\caption{{\it Ab initio} calculated values of $\mathbf{e}$ and $\mathbf{d}$ for both \hfo\ and \zro.}
\begin{tabular}{|c|ccc|ccc|}
\hline
      & \multicolumn{3}{c|}{\zro}                                                        & \multicolumn{3}{c|}{\hfo}                                                        \\ \hline
      & \multicolumn{1}{c|}{$\mathbf{\overline{e}}$} & \multicolumn{1}{c|}{$\mathbf{e}$} & $\mathbf{d}$ & \multicolumn{1}{c|}{$\mathbf{\overline{e}}$} & \multicolumn{1}{c|}{$\mathbf{e}$} & $\mathbf{d}$ \\
Index & \multicolumn{1}{c|}{[C/m$^2$]} & \multicolumn{1}{c|}{[C/m$^2$]} & [pm/V] & \multicolumn{1}{c|}{[C/m$^2$]} & \multicolumn{1}{c|}{[C/m$^2$]} & [pm/V] \\ \hline
31    & \multicolumn{1}{c|}{-0.40}                   & \multicolumn{1}{c|}{-1.78}        & -3.03        & \multicolumn{1}{c|}{-0.37}                   & \multicolumn{1}{c|}{-1.40}        & -1.84        \\ \hline
32    & \multicolumn{1}{c|}{-0.37}                   & \multicolumn{1}{c|}{-1.54}        & -2.17        & \multicolumn{1}{c|}{-0.39}                   & \multicolumn{1}{c|}{-1.53}        & -2.51        \\ \hline
33    & \multicolumn{1}{c|}{0.69}                    & \multicolumn{1}{c|}{-1.56}        & -2.56        & \multicolumn{1}{c|}{0.65}                    & \multicolumn{1}{c|}{-1.34}        & -2.03        \\ \hline
15    & \multicolumn{1}{c|}{-0.32}                   & \multicolumn{1}{c|}{-0.26}        & -2.97        & \multicolumn{1}{c|}{-0.29}                   & \multicolumn{1}{c|}{-0.22}        & -2.39        \\ \hline
24    & \multicolumn{1}{c|}{-0.24}                   & \multicolumn{1}{c|}{0.78}         & 9.45         & \multicolumn{1}{c|}{-0.22}                   & \multicolumn{1}{c|}{0.69}         & 7.46         \\ \hline
\end{tabular}
\label{tab3}
\end{table}

\begin{figure}[h!]
\centering
\includegraphics[width=9.0cm]{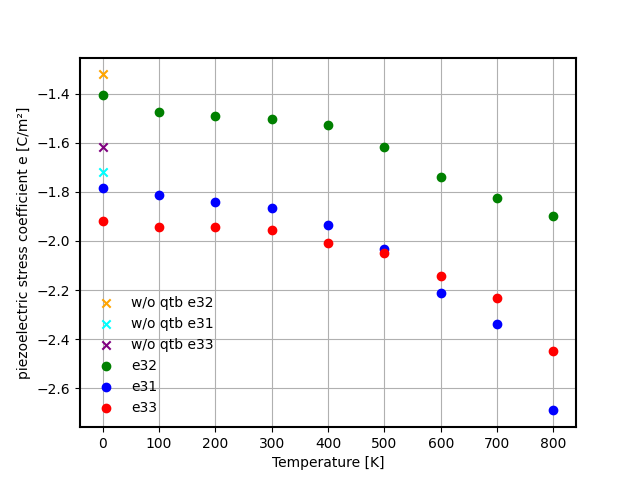}
\caption{Piezoelectric stress coefficients obtained by unidirectionally straining the po-phase cell. Above 900K the applied strain induces the irreversible decay from po- into t-phase.}
\label{fig:stresspiezo}
\end{figure}

From the MD simulations and $\pm 1 \%$ unidirectional strain amplitudes in $0.5 \%$ steps we calculated the polarization derivative using central differences. The results in Fig. \ref{fig:stresspiezo} show values slightly larger than from ab initio. For example for $e_{33}$ we obtain -1.9C/m$^{2}$ compared to an ab initio value of -1.56C/m$^{2}$. 
Furthermore the absolute values increase with temperature beyond room temperature.
To explain the trend of the MD calculated piezoelectric coefficients being larger than from DFT calculations, we performed near $T=0K$ simulations without the QTB model. The results in Fig. \ref{fig:stresspiezo} show smaller values being close to the DFT results. The QTB model therefore reduces the temperature dependence of the piezoelectric coefficients below 300K.

The increase of the absolute values of the piezoelectric stress coefficients towards the phase transition temperature $T_1$ indicates destabilization of the po-phase. In MD simulation we cannot approach $T_1$ closely because the tensile strained po-phase starts already to decay at 900K into the t-phase. This is a simulation effect because we need finite strain perturbation which increases the instability.

\subsection{\label{sec:  Piezoelectric strain coefficient}Piezoelectric strain coefficients}

\begin{figure}[h!]
\centering
\includegraphics[width=9.0cm]{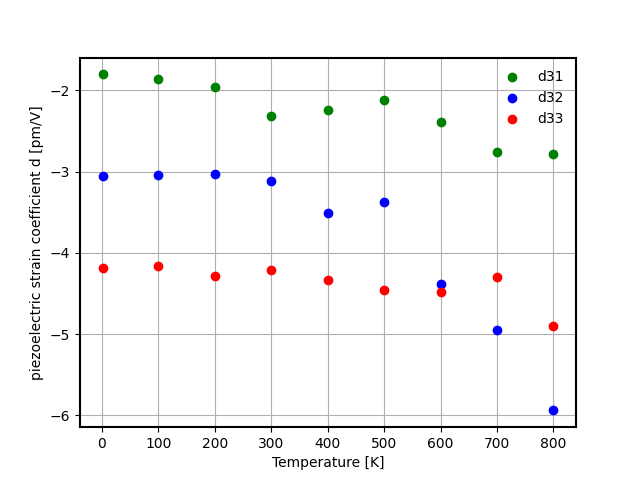}
\caption{Piezoelectric strain coefficients, obtained by unidirectionally stressing the po-cell. Above 900K a irreversible phase-transition from the po- to the t-phase occurs.}
\label{fig:strainpiezo}
\end{figure}

The piezoelectric strain coefficients describe the change of polarization under controlled stress $\sigma_j$ and zero electric field $\cal{E}$
\begin{equation}
    d_{\alpha j} = \left.\frac{\partial P_\alpha}{\partial \sigma_j}\right\vert_{\cal{E}}
                 = \left.\frac{\partial \eta_j}{\partial {\cal{E}}_{\alpha}}\right\vert_{\sigma}
\end{equation}
The coefficients can be obtained according to the first relation by applying stress to the crystal and calculating the response of the polarization, which requires a NPT simulation with a barostat to fix the stress. Using a barostat the error of the resulting polarization becomes larger than for the polarization calculated with a fixed strain. Fig.   \ref{fig:strainpiezo} shows the results as a function of temperature. The values are negative between -2pm/V and -4pm/V and increase with temperature. They are slightly larger than the ab initio results in Table \ref{tab3}, which range from -2pm/V to -3pm/V. Again, this is an effect of the QTB model.
The values increase with temperature and become very large when the phase transition temperature $T_1$ is approached. Stable values very close to $T_1$ cannot be obtained because the finite stress perturbation furthermore destabilizes the po-phase.
The phase transition at $T_1$ changes the polarization about a very large negative value to zero. But this very large change cannot be interpreted as giant, negative strain coefficient because the transition is irreversible. In contrast, the giant piezoelectric strain coefficient section \ref{sec:giant}
has the t-phase as initial state, is positive and is reversible beyond $T_1$.

Thin film \zro\ often consists of a phase mixture of t-phase, po-phase and eventually m-phase grains. Whereas the sign of the piezoelectric coefficient of a single grain depends on the crystal phase, the averaged piezoelectric coefficient of a thin film depends on the phase mixture. For a thin film it is probably difficult to measure a negative piezoelectric coefficient, because this requires a nearly pure po-phase film.

\subsection{\label{sec: Pyroelectric coefficient}Pyroelectric coefficient}
\begin{figure}[h!]
\centering
\includegraphics[width=9.0cm]{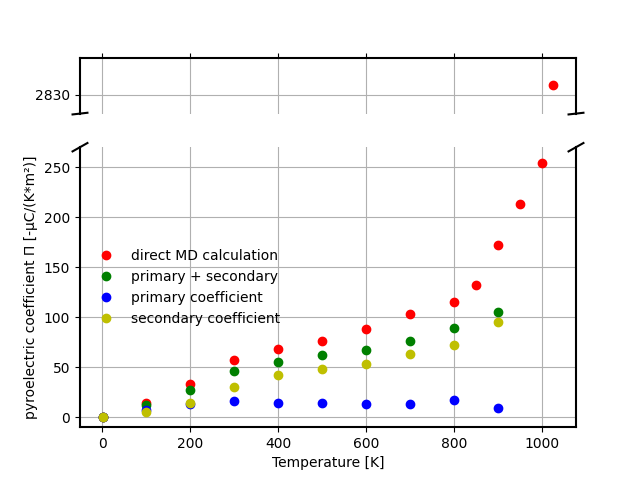}
\caption{Pyroelectric coefficient of \zro\ calculated as sum of primary and secondary contribution, as well as calculated in a single simulation without limiting assumptions. The very large value is obtained from polarization change after phase transition beyond $T_1$.}
\label{fig:pyro}
\end{figure}

In ab initio calculations, the pyroelectric coefficients $\Pi_{\alpha}$, $\alpha=x,y,z$, are separated into the primary contribution $\Pi^{(1)}$ describing the change of polarization at fixed cell parameters and variable temperatures, and the secondary contribution $\Pi^{(2)}$ which describes the effect from thermal expansion of the cell together with the change of polarization via the piezoelectric effect.\cite{Bernardini1997} 
\\
\begin{equation}
    \Pi_{\alpha} = \Pi_{\alpha}^{(1)} + \Pi_{\alpha}^{(2)} = \left(\frac{\delta P_{\alpha}}{\delta T}\right)_\eta + \sum_j \left(\frac{\delta P_{\alpha}}{\delta \eta_j}\right)_T \left(\frac{\delta \eta_j}{\delta T}\right)_\sigma
    \label{eq:decomposition}
\end{equation} 
$\Pi^{(1)}$ and $\Pi^{(2)}$ have been calculated for \hfo\ from ab initio by Liu\cite{liu}. We calculated the $\Pi^{(1)}$ with NVT simulations for different temperatures and fixed cell parameters from 0K. This leads to a change in polarization attributed to internal, temperature-dependent rearrangements of the ions.
The secondary pyroelectric coefficient $\Pi^{(2)}$ was obtained as product of the previously from MD calculated thermal expansion coefficients and piezoelectric stress coefficients. Due to the negative sign of the piezoelectric coefficient, the thermal expansion of the crystal leads to a decrease of the polarization. The peculiarity of \zro\ consists in the negative signs of primary and secondary contribution leading to an added effect in total. Both contributions as well as the sum are shown in Fig. \ref{fig:pyro}. The increase of the values with temperature is moderate.

In MD a NPT simulation yields the pyroelectric coefficients directly, without any limiting assumptions
 \begin{equation}
    \Pi_{\alpha} = \left(\frac{dP_{\alpha}}{dT}\right)_\sigma 
\end{equation} 

The resulting values shown in Fig. \ref{fig:pyro} start at zero which is a result of the quantum occupation at low temperature, contained in the QTB model. The values for \zro\ are somewhat larger than reported for \hfo\cite{liu} and increase significantly beyond these values when the phase transition temperature is approached.
The direct calculation shows larger values than from the decomposition. This hints to the assumptions made in deriving eq. \ref{eq:decomposition} as truncated part of some expansion. The simulated large values derive from anharmonic effects and the destabilization of the po-phase.
The significant increase of the pyroelectric coefficient when approaching $T_1$ is consistent with the large values which have been found experimentally\cite{mart}.

Fig. \ref{fig:pyro} shows furthermore a negative, giant value obtained beyond $T_1$ after the phase transition has occurred. This hints to values which can be obtained when the phase transition can be reversed. In the experiments of Hoffmann et al.\cite{Hoffmann-pyro} the polarization change of the po-phase was observed for temperatures around and beyond $T_1$, and negative, giant values for the pyroelectric coefficients could be derived. But in the experiments the po-phase was recovered with help of electric field-cycling, which spontaneously does not happen beyond $T_1$. Therefore this giant pyroelectric coefficients cannot be compared with the very large pyroelectric coefficients obtained below $T_1$.

\begin{figure}[h!]
\centering
\includegraphics[width=9.0cm]{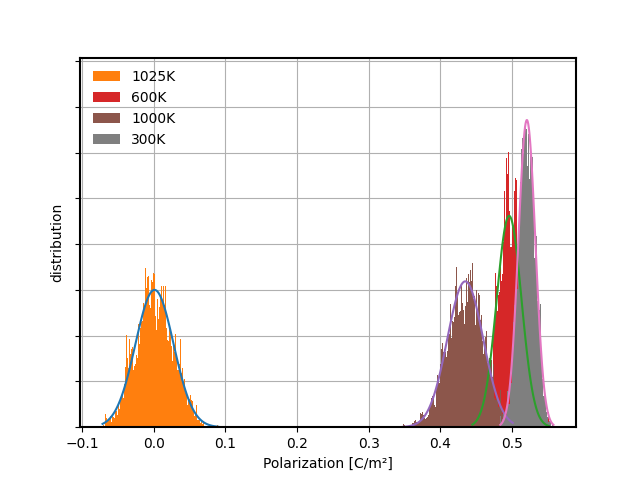}
\caption{The distributions show the spread of polarization in our supercell at different temperatures. Increasing the temperature leads to a broadening of the polarization distribution. The po-phase decays exhaustive to the t-phase.}
\label{fig:histogram}
\end{figure}

To get a picture of the polarization change close to the phase-transition, the polarization distribution at different temperatures is plotted, see Fig. \ref{fig:histogram}. The polarization decreases between 0K and 1000K from 0.54C/m$^{2}$ to 0.43C/m$^{2}$, which is a reduction of merely 20\%.
 
%In the range up to 700K this equation remains true simulations yield the primary pyroelectric coefficient and the piezocoefficient multiplied by the %thermal expansion yield the secondary pyroelectric coefficient, which add up to the total pyroelectric coefficient. Those values lay in the margin of %error with the whole pyroelectric coefficient obtained from, zero stress, NPT simulations. 
%\\

\subsection{Dielectric constant}

\begin{figure}[h!]
\centering
\includegraphics[width=9.0cm]{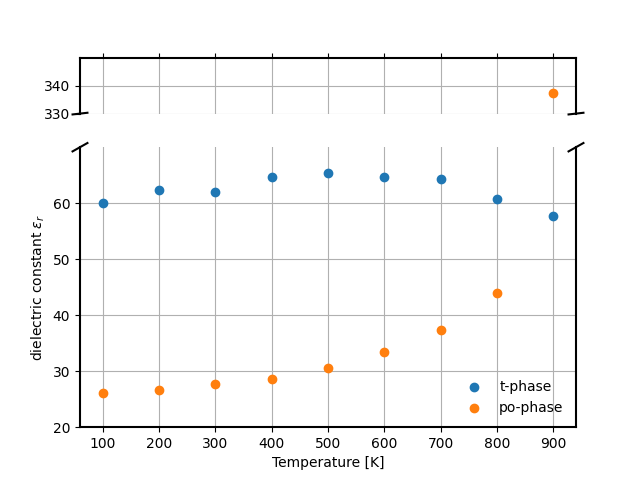}
\caption{Relative permitivity $\epsilon_{33}$ of the po- and the t-phase from NPT MD simulation. For the po-phase the value increases significantly close to $T_1$.}
\label{fig:diel}
\end{figure}

We performed NPT simulations at different temperatures, applied electrical field perturbation, and obtained the temperature dependent dielectric constant. Fig. \ref{fig:diel} shows the results for the t-phase and the po-phase. The dielectric constant of the t-phase matches with $\epsilon_{33}$=60 ab initio data \cite{Fischer} and experimental values very well and shows no temperature dependence, again demonstrating the surprising stability at $T_1$. The dielectric constant of the po-phase, however, shows a low temperature value of $\epsilon_{33}$=26 and a significant increase with temperature leading to a very large value when $T_1$ is approached. Very large values for $\epsilon$ have recently been measured for a thin film with HZO in predominatly polar phase, when the phase transition temperature $T_1$ was approached\cite{Schroeder2022}.

\section{\label{sec:Conclusion}Conclusion}

In summary, we applied molecular dynamics with a machine learned potential to identify the underlying mechanisms for piezoelectric, pyroelectric and dielectric effects in monocrystalline \zro. 
Another motivation was to find an estimate of what values to expect for these coefficients and what circumstances would be required.
The developed potential model reproduces and predicts ab initio data very well, but involves their imperfection.

We found the phase transition temperature $T_1$ to govern the overall behaviour. This temperature differs from the Curie-temperature and results from the inherent kinetic energy barriers between the po-phase and the t-phase. The barrier depends in simulation quantitatively on the chosen density functional.
We could distinguish the genuine piezoelectric effects with negative coefficients increasing towards $T_1$ which requires the po-phase, from the giant piezoelectric strain effect. The latter is positive, involves a field-induced phase transition and requires the t-phase. The measured large positive\cite{kirbach-piezo} and negative\cite{mart} values can be attributed to these effects. In a phase mixed polycrystalline thin film effects of both sign should coexist and the positive effect should mostly dominate.
The largest possible piezoelectric coefficients would result from the giant effect under the conditions of maximal t-phase content. A further optimization would result from maximizing the volume change\cite{falkowski} and simultaneously minimizing the kinetic barrier for the required electric field, possibly with doping.

The direct simulation of the pyroelectric coefficient compared to the simulation of the primary and secondary contribution revealed, that the enhancement of the effect with temperature is related to anharmonic effects not contained in the defining truncated expression. The simulated pyroelectric coefficients increase significantly towards the phase transition temperature $T_1$ which fits well to the experimentally observed values.
The largest pyroelectric coefficients should be achieved with a film of largest po-content, and a temperature closest to $T_1$. Further effects from doping which increase the anharmonic contributions have already been demonstrated exemplarily by Liu\cite{liu}.
Crossing the phase transition temperature $T_1$ does not lead to the giant pyroelectric coefficient, because the po-phase does not recover without an action. An electric field is required to complete the recovery of the po-phase, which matches the circumstances of the observation of giant piezoelectricity by Hoffmann\cite{Hoffmann2015}.

Regarding the dielectric constant, we found no dielectric enhancement for the t-phase approaching the phase transition temperature, which convincingly demonstrates the stability of the t-phase, and leaves us with the puzzle of the po-phase formation from a tetragonal precursor phase. The po-phase, however, showed a significant dielectric enhancement, close to a singularity at $T_1$. This surprising observation has been done recently in experiments in \hzo\ and is expected for a proper ferroelectric material like the classical BaTiO$_3$ ferroelectric. The explanation from our simulation is an energy landscape with a barrier for the transition from t-phase to po-phase below $T_1$, but no barrier for the transition from po-phase to t-phase above $T_1$.

It is expected that the insights gained from the simulations with \zro\ can be transferred to \hfo\ and the doped materials. Most important is here the polar to tetragonal phase transition temperature $T_1$, and the phase mixture. Further effects on the phase transition temperature are expected from size effects resulting from the grains in polycristalline films.

\begin{acknowledgments}
Luis Azevedo Antunes and Richard Ganser received funding from the Deutsche Forschungsgemeinschaft (German Research Foundation) in the frame of the project "Zeppelin" (Project KE 1665/5-1). The authors gratefully acknowledge the Leibniz Supercomputing Centre for funding this project by providing computing time on its Linux-Cluster.
\end{acknowledgments}

%\nocite{*}
\bibliography{article}% Produces the bibliography via BibTeX.
% Please add the following required packages to your document preamble:
% \usepackage[table,xcdraw]{xcolor}
% If you use beamer only pass "xcolor=table" option, i.e. \documentclass[xcolor=table]{beamer}
% Please add the following required packages to your document preamble:
% \usepackage[table,xcdraw]{xcolor}
% If you use beamer only pass "xcolor=table" option, i.e. \documentclass[xcolor=table]{beamer}

\end{document}

\title{\emph{Supplementary Material}\\Piezo- and pyroelectricity in Zirconia:  a study with machine learned force fields}

\author{Richard Ganser}%
\affiliation{Department of Applied Sciences and Mechatronics, Munich University of Applied Sciences, Lothstr. 34, 80335 Munich, Germany}%
\affiliation{These two authors contributed equally to this work.}

\author{Simon Bongarz}
\affiliation{Department of Applied Sciences and Mechatronics, Munich University of Applied Sciences, Lothstr. 34, 80335 Munich, Germany}%
\affiliation{These two authors contributed equally to this work.}

\author{Alexander von Mach}%
\affiliation{Department of Applied Sciences and Mechatronics, Munich University of Applied Sciences, Lothstr. 34, 80335 Munich, Germany}%

\author{Luis Azevedo Antunes}%
\affiliation{Department of Applied Sciences and Mechatronics, Munich University of Applied Sciences, Lothstr. 34, 80335 Munich, Germany}%

\author{Alfred Kersch}
\email{alfred.kersch@hm.edu}
\affiliation{Department of Applied Sciences and Mechatronics, Munich University of Applied Sciences, Lothstr. 34, 80335 Munich, Germany}%

\date{\today}% It is always \today, today,
             %  but any date may be explicitly specified

\maketitle
\section*{S1: Structure, elastic constants, piezo constants}
The 12-atomic structures of \zro\ and \hfo\ were calculated  with ABINIT\cite{abinit} using PBEsol\cite{pbesol} with norm conserving pseudopotentials from the PseudoDojo library\cite{pseudodojo},
$6 \times 6 \times 6$ k-point grids, and an energy cutoff of 80Ha and 100Ha, respectively.

\begin{table}[h!]
\centering
 \caption{Calculated lattice constants and Wyckhoff positions for \zro\ and \hfo.}
  \begin{tabular}{@{}rcccc}
  \hline
  \multicolumn{5}{c}{ZrO$_2$\qquad a=5.050[\AA], b=5.255[\AA], c=5.069[\AA]} \\
    \hline
    Atom & Wyckhoff Pos. & x & y & z \\
    \hline
Zr  & 4a &  0.03119 &  0.23282 & -0.00256 \\
OI  & 4a &  0.36709 &  0.42996 & -0.14130 \\
OII & 4a &  0.26799 & -0.03806 &  0.25024 \\
\hline
  \multicolumn{5}{c}{HfO$_2$\qquad a=5.013[\AA], b=5.223[\AA], c=5.037[\AA]} \\
\hline
Hf  & 4a &  0.03274 &  0.23249 & -0.00248 \\
OI  & 4a &  0.36789 &  0.42836 & -0.14025 \\
OII & 4a &  0.26616 & -0.04016 &  0.24910 \\
\hline
  \label{tab1}
  \end{tabular}
\end{table}

The elastic constants and the piezoelectric constants were calculated with the density function perturbation theory, as implemented in Abinit.
The elastic tensor with relaxed atoms results from the combination of the second derivative for internal stress and the constants of the interatomic
force; while the piezoelectric tensor with relaxed atoms takes into account the second derivatives for the internal stress, interatomic force constants, and Born effective charges.

\begin{table}[h!]
\centering
 \caption{Elastic constants for \zro\ calculated from Abinit (light) in comparison with the constants calculated from the DP model (bold).}
  \begin{tabular}{@{}rrrrrrrrr}
  \hline
                  \multicolumn{9}{c}{elastic constants ZrO$_2$ [GPa]}  \\
    \hline
    C_{11} & C_{22} & C_{33} &  C_{12} & C_{13} & C_{23} &  C_{44} & C_{55} & C_{66} \\
      377  &   370  &   364  &   147   &   126  &   113  &     86  &    83  &   122  \\
      $\mathbf{347}$&$\mathbf{370}$&$\mathbf{351}$&$\mathbf{138}$&$\mathbf{94}$&$\mathbf{102}$&$\mathbf{83}$&$\mathbf{102}$&$\mathbf{126}$\\
      \hline
  \label{tab1}
  \end{tabular}
\end{table}

\newpage
\section*{S2: Phonon dispersion}
The dynamical stability of the structures was checked with a finite displacement phonon calculation using phonopy\cite{Togo2015} within the harmonic approximation (HA), using 2 x 2 x 2 supercells. The Brillouin zone samping was conducted with a 10 x 10 x 10 q mesh. The forces were calculated with DFT using FHI-aims, and LAMMPS using the DP model.
\begin{figure}[h!]{\columnwidth}
\centering
\includegraphics[width=9.0cm]{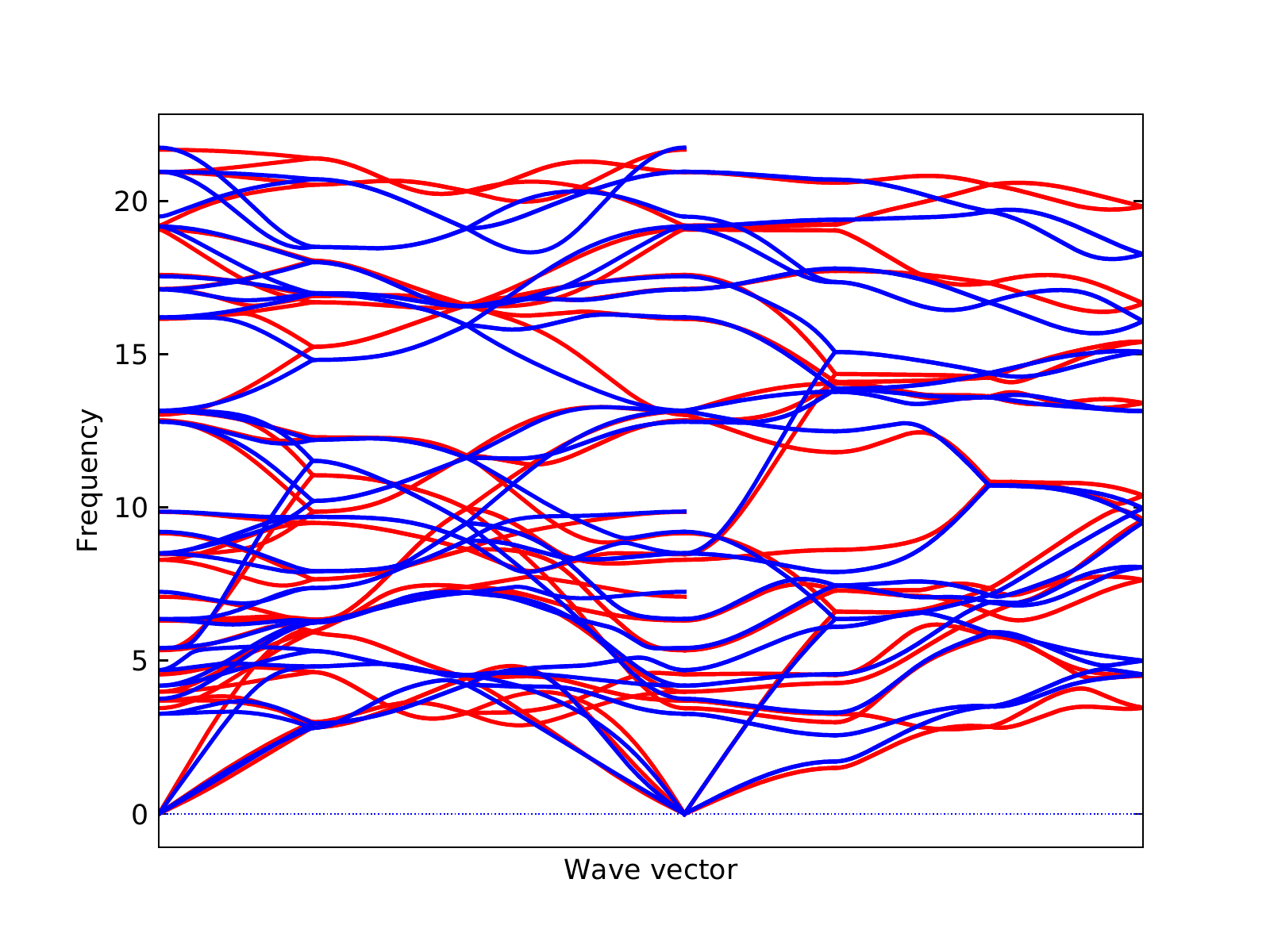}
\caption{Phonon-dispersion relation of the no 137 calculated with DFT (blue) and DP (red).}
\label{fig:NEB}
\end{figure}
\begin{figure}[h!]{\columnwidth}
\centering
\includegraphics[width=9.0cm]{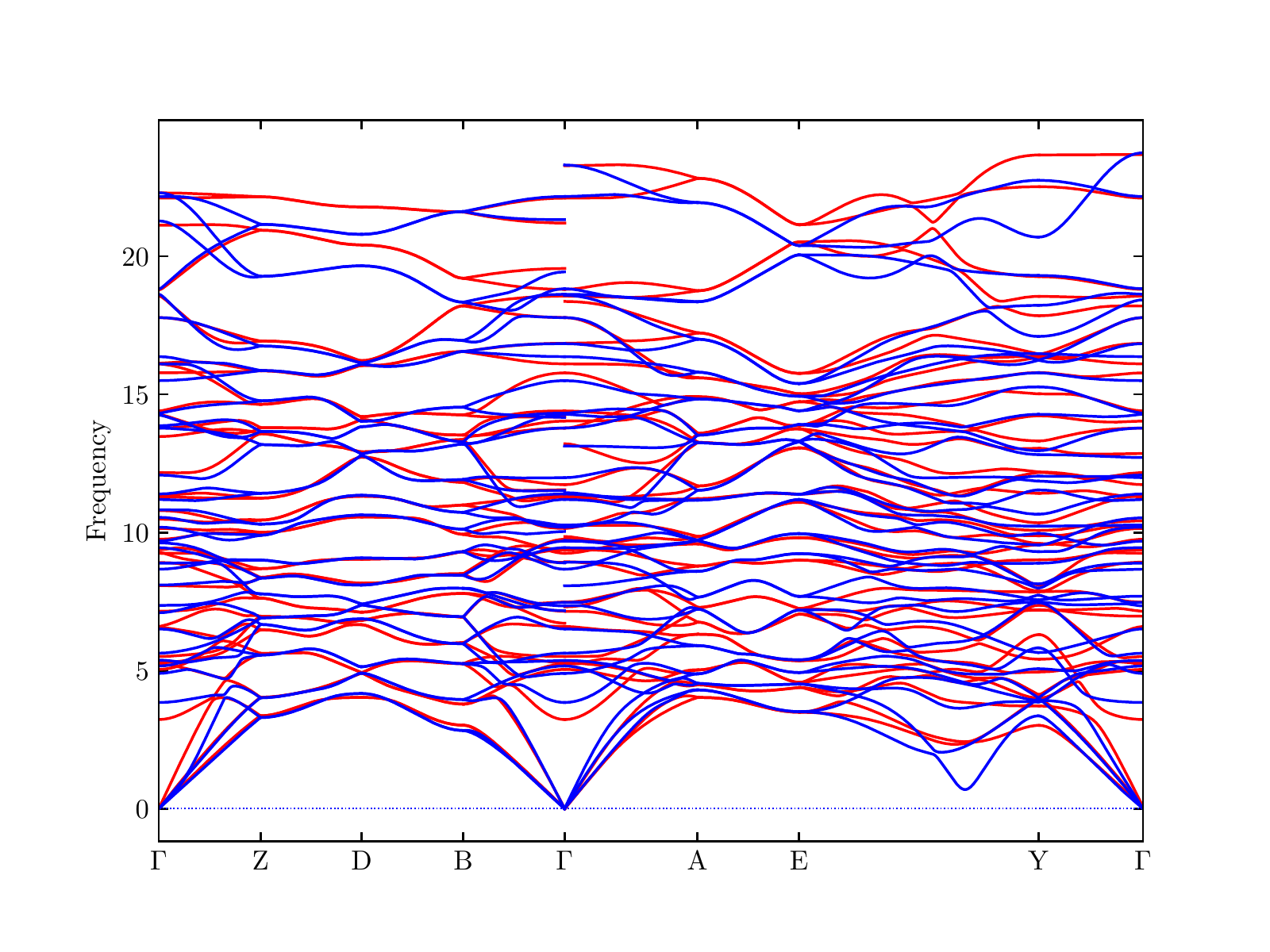}
\caption{Phonon-dispersion relation of the no 14 calculated with DFT (blue) and DP (red).}
\label{fig:NEB}
\end{figure}
\begin{figure}[h!]{\columnwidth}
\centering
\includegraphics[width=9.0cm]{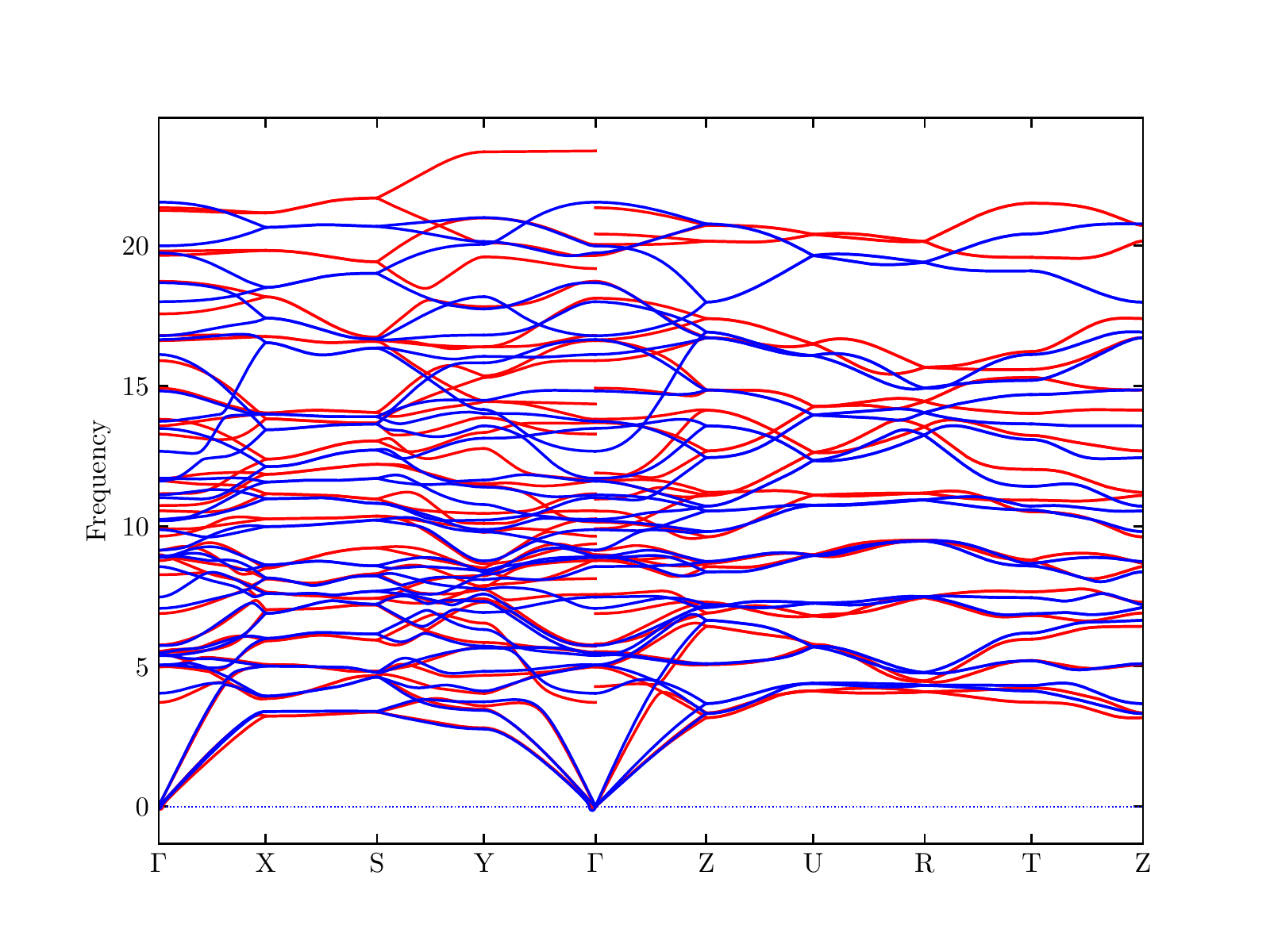}
\caption{Phonon-dispersion relation of the no 29 calculated with DFT (blue) and DP (red).}
\label{fig:NEB}
\end{figure}

\newpage

\section*{S3: Thermal expansion}

To validate the thermal expansion coefficients with experimental data we calculated the coefficients for the t-phase and the m- phase using NPT ensembles.

\begin{figure}[h!]{\columnwidth}
\centering
\includegraphics[width=9.0cm]{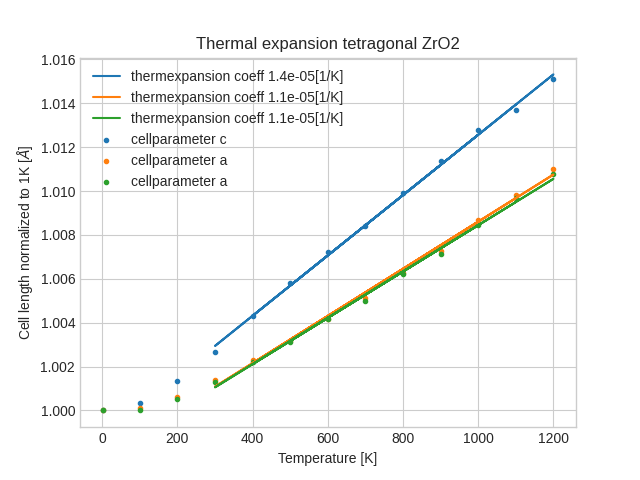}
\caption{Complete and irreversible phase-transition between the polar-orthorhombic and the tetragonal phase.}
\label{fig:NEB}
\end{figure}
\begin{figure}[h!]{\columnwidth}
\centering
\includegraphics[width=9.0cm]{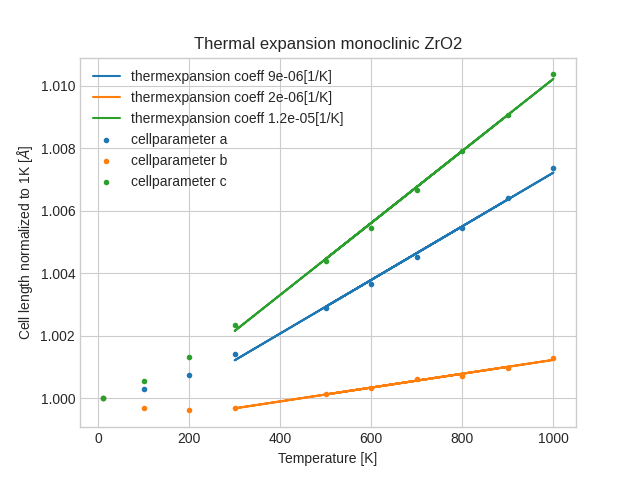}
\caption{Complete and irreversible phase-transition between the polar-orthorhombic and the tetragonal phase.}
\label{fig:NEB}
\end{figure}

\begin{table}[h!]
\centering
 \caption{Comparison of experimental and DP model calculated axial thermal expansion coefficient in $\times 10^6\ 1/K$.}
  \begin{tabular}{@{}c c c c l}
\hline
\multicolumn{5}{c}{monoclinic \zro}\\
a   &  b  &   c    & &  reference \\
\hline
10.3 & 1.35 &  14.7 & &  Patil\cite{Patil}  around  $600C^{\circ}$\\
 8.4 & 2.9  &  11.2 & &  Haggerty\cite{Haggerty2014}   around  $600C^{\circ}$\\
 9   & 2    &  12   & &  this work DP around  $600C^{\circ}$\\
\hline
\multicolumn{5}{c}{tetragonal \zro}\\
a   &  b  &   c    & &  reference \\
\hline
11.6 & - &   16.1   & &  Patil\cite{Patil}  around  $1200C^{\circ}$\\
10.9 & - &   16.8   & &  Haggerty\cite{Haggerty2014}   around  $900C^{\circ}$\\
11   & - &   14     & &  this work DP around  $900C^{\circ}$\\
\hline
\end{tabular}
\end{table}

\section*{References}

\bibliography{supplementary}